\documentstyle[prd,aps,floats]{revtex}
\begin{document}
\draft

\def\nc{noncommutative }
\def\ncy{noncommutativity }
\def\com{commutative }
\def\ncs{\Lambda_{{\rm NC}}}
\def \simlt{\stackrel{<}{{}_\sim}}
\def \simgt{\stackrel{>}{{}_\sim}}
\newcommand{\be}{\begin{equation}}
\newcommand{\ee}{\end{equation}}
\newcommand{\bea}{\begin{eqnarray}}
\newcommand{\eea}{\end{eqnarray}}
\newcommand{\ba}{\begin{array}}
\newcommand{\ea}{\end{array}}   

\input epsf \renewcommand{\topfraction}{0.8} 
\twocolumn[\hsize\textwidth\columnwidth\hsize\csname 
@twocolumnfalse\endcsname

\title{Noncommutativity in Space and Primordial Magnetic Field}

\author{Anupam Mazumdar and Mohammad M. Sheikh-Jabbari}

\address{The Abdus Salam International Center for Theoretical
Physics, Strada Costiera 11, Trieste, Italy} 
\date{\today} 
\maketitle
\begin{abstract} 
In this paper we show that noncommutativity in spatial coordinates can generate
magnetic field in the early Universe on a horizon scale. The strength of 
such a magnetic field depends on the number density of massive charged 
particles present at a given moment. This allows us to trace back the
temperature dependence of the noncommutativity scale from the bounds on
primordial magnetic field coming from nucleosynthesis.
\end{abstract}

\pacs{PACS: 98.80.Cq, 11.25.-w, 11.25.Sq. \hspace{6cm} hep-ph/0012363}
 
\vskip 1cm]


Although the idea of having a \nc space-time is an old one \cite{Sny},
until a renewed motivation from string theory \cite{SW}, it has not been
studied seriously. The operators corresponding to coordinates of a \nc
space (Moyal plane), ${\hat{x}}^{\mu}$, satisfy
\begin{equation}\label{cr}
[{\hat x}^{\mu},{\hat{x}}^{\nu}]=i\theta^{\mu\nu}\, ,
\end{equation}
where $\theta^{\mu\nu}$ is a given constant which has  dimension of [length]$^2$.
For our purposes it is more convenient to introduce the \ncy scale, $\Lambda_{{\rm
NC}}$, given by 
\be
\theta^{\mu\nu}={c^{\mu\nu}\over \Lambda_{{\rm NC}}^2}\ ,
\ee
where $c^{\mu\nu}$ is an antisymmetric tensor whose components are ${\cal O}(1)$.
In order to obtain the \nc version of a given field theory generally 
one can use the following prescription. Take the classical action for the field 
theory and replace the product of the fields by a $\star$-product. Such as
\begin{eqnarray}
\label{star}
(f\star g)(x)&=&\exp{\bigl({i\over 2}\theta_{\mu\nu}
\partial_{x_{\mu}}\partial_{y_{\nu}}\bigr)}~f(x)g(y)\Big|_{x=y}\cr
&=&f(x)g(x)+{i\over 2}\theta_{\mu\nu}\partial_{{\mu}}f\ \partial_{{\nu}}g+{\cal
O}({\theta^2})\,.
\end{eqnarray}
We notice the $\star$-product in Eq.~(\ref{star}) is a \nc one, 
$f\star g\neq g\star f$. For some useful relations on the 
$\star$-product calculus, see Ref.~\cite{Andrei}. It has been noticed that there 
are problems with the unitarity and causality in the \nc space-time if 
$\theta_{0i}\neq 0$, \cite{unitray}. Therefore, in this paper  
we restrict ourselves to the \nc space, i.e. $\theta_{0i}=0$.

Studying various physical consequences of such \ncy in spatial coordinates is 
of great interest. In this regard both quantum mechanical systems and 
field theory results can be used to constrain the lower bound on the 
\ncy scale. For the \nc quantum mechanics, the hydrogen
atom in a \nc space has been discussed in Ref.~\cite{Lamb}, and its spectrum
has also been calculated up to first order in $\theta$. There it is shown 
that \ncy lifts the degeneracy of some states in the spectrum 
and in particular it changes the Lamb-shift. For a general quantum 
mechanical system with an electro-magnetic interactions the 
Hamiltonian receives a correction due to \ncy \cite{Lamb}.
Such a correction at tree level up to first order in $\theta$ is given by
assigning a momentum dependent electric dipole moment; $\vec{d_e}$ \cite{{Ihab}} 
\be
\label{dip}
d_e^i={e\over 2\hbar}\theta^{ij}p_{j} \,.
\ee

The \nc field theories have their own attractions. The very basic 
question of the renormalizability of these field theories have been 
discussed in many papers, for some relevant references, see Ref.~\cite{Martin}.
It has been shown that \nc version of a real $\phi^4$ theory is 
renormalizable up to two loops \cite{Andrei,Aref,Sei}. The 
noncommutative QED (NCQED) has also been discussed in Ref.~\cite{Ihab,Haya}, and shown 
to be renormalizable at one loop level. It was noticed that unlike 
the usual QED, NCQED is an asymptotic free theory.
However, $\beta$-function {\it does not} depend on $\theta$ 
\cite{Haya}\footnote{In general $\theta\to 0$ limit is not a 
smooth one \cite{Andrei}.}.

By treating the \nc space as a natural extension to the usual space,
many authors have studied the phenomenological consequences. By
considering various $2\to 2$ scattering processes in NCQED, and, 
comparing the cross sections with that of the usual standard model, 
some lower bounds on the \ncy scale has been obtained 
in Refs.~\cite{Slac,NLC}. Here, we would like to mention the crucial point 
that in the \nc models the Lorentz invariance is explicitly broken. As
a result of that all the amplitudes are frame dependent.
According to the results obtained from studying the cross-sections,
the lower bound on \ncy has been confirmed to be around  1000-2000 Gev \cite{Slac}. 
Besides the scattering processes, the hydrogen atom spectrum and the Lamb-shift
can also be used to put some bounds on $\theta$, which leads to $\ncs \simgt 10^4$
Gev \cite{Lamb}. One can still obtain better bounds on $\ncs$ up to $10^5$ Gev from
the neutron electric dipole moment \cite{Goran}. 

As an interesting property of \nc space, it has been noticed in Ref.~\cite{Ihab}
that a magnetic dipole moment of a charged massive particle (e.g. electron) 
at one loop level receives some quantum corrections, which unlike the usual 
\com case is spin {\it independent}, and, it is directly proportional to $\theta$. 
Now, this can have an interesting consequence at a macroscopic level and a direct
implication in the early Universe, such as generating magnetic fields at a cosmological
scale. As we shall see the primordial magnetic fields constrains $\theta$, and
this may improve the lower bound on \ncy scale.
As mentioned earlier, in the \nc space the Lorentz symmetry is violated, and in fact
this magnetic field is also frame dependent. Our setup is good for a very slowly moving
frame, i.e. what we actually have on the Earth, in which the isotropic cosmos
assumption is a valid one. 

As discussed earlier in Ref.~\cite{Ihab}, in NCQED the magnetic dipole moment 
of \nc Dirac particle at one loop is given by
\be\label{magnet}
\langle \vec \mu \rangle =\frac{eq}{2m}g \vec S +
\frac{e\alpha \gamma_{e}}{6 \pi} q^3 m\, \vec\theta\ ,
\ee
where $\gamma_{\rm e}\sim 0.57$ denotes the Euler number,
$m$ is the mass of a particle which carries charges $eq$,
$\vec{S}$ is spin of the particle, vector $\vec{\theta}$ is defined as
$\theta_i=\epsilon_{ijk}~\theta^{ij}$, and, the gyromagnetic factor 
is denoted by $g$, which at one loop is given by
$$
g=2+{\alpha~q^2\over \pi},\;\;\;\;\;\;\;\ \alpha={e^2\over
4\pi}={1\over 137}\ ,
$$
The first term in Eq.~(\ref{magnet}) is the usual (commutative)
expression, while the
second term is due to noncommutativity and it is zero for a massless particle.
Another important fact about the second term in Eq.~(\ref{magnet}) is
{\it unlike the first term}, it is {\it invariant} under C (charge conjugation) 
and CP. This can be understood by  noticing that under C (and CP) \cite{CPT} 
$$ 
\theta \stackrel{{\rm C}}{\longrightarrow} -\theta\ ,\ \ \ \,
\theta \stackrel{{\rm CP}}{\longrightarrow} -\theta\ .
$$

Now, in order to obtain a large scale magnetic field originating from the 
matter content in a given volume, we must sum over all the possible 
states of the particles as
\be
\label{imp0}
\vec{B}=\sum_{i,\vec{S}}\ \langle \vec \mu_{i,\vec{S}} \rangle\ n_i\ ,
\ee
where $i$ runs over all the ``relevant set'' of particles, where $n_i$ is the
corresponding number density. The above expression for the magnetic field is 
quite interesting and as we have mentioned earlier, such a magnetic field
spread in a large scale can be originated very easily in the early Universe. 
Especially, there exists large scale magnetic fields
in galaxies and galaxy clusters with a large coherence length and a 
typical strength of $10^{-6}~{\rm G}$, at a redshift of $z=0.395$ \cite{kronberg}. 
It is usually assumed that there must be a galactic dynamo mechanism which 
can amplify the pre-existing primordial magnetic field. A recent study in 
this direction suggests that a weak magnetic field of strength ${\rm B}_{\rm g} \sim 
10^{-30}$G during the galaxy formation can be sufficient to amplify the strength
up to the present limit, provided there is a non-vanishing cosmological constant
\cite{lilley}. By now, there are many mechanisms which claims to understand 
the existence of the primordial magnetic field, acting as a seed. For a nice review, 
we refer the readers to Ref.~\cite{grasso}. 

However, the magnetic field due to the presence of \ncy has a marked difference.
First of all the generation of this magnetic field is continuous and accumulative
in nature. All it demands is the presence of charged massive particles present in 
any corner of the Universe. For such relativistic species 
the number density behaves as $n \approx {\rm T}^3$, where $T$ is the 
temperature of the Universe.
Thus we see that the strength of such a magnetic field depends on the temperature 
of the Universe. At this stage one might wonder what could be the strength of 
such a field at electroweak scale and at nucleosynthesis scale. 
It is a well known fact that
a strong magnetic field can alter thermodynamical distribution of the 
charged particles by modifying the phase space of volumes of the particles and the 
antiparticles, which can also affect the weak interaction processes.
It was first noticed in Ref.~\cite{matese1}, that a strong magnetic 
field can significantly enhance the $\beta$ decay rate of the neutrons. On
this basis the authors have argued that relic abundance for 
$^{4}{\rm He}$ will have a strong suppression compared to the standard case
\cite{matese2}, because below the freezing temperature for the neutrons and protons,
there is a finite time before they can coalesce to form a composite nuclei.
In that finite time if the neutron abundance decreases due to its decay, it 
is natural that not only $^{4}{\rm He}$ abundance is affected but also 
the abundances of heavier elements. Based on these facts 
the authors in Ref.~\cite{dario} have derived constraints on the primordial 
magnetic field at the beginning of nucleosynthesis, corresponding to a 
temperature ${\rm T} \sim 10 {\rm MeV}$. 
${\rm B}({\rm T}=10 {\rm MeV})=10^{-8}{\rm GeV}^2 ~~{\rm for}
~{\rm Y}_{\rm p} =0.236$, where ${\rm Y}_{\rm p}$ denotes $^{4}{\rm He}$ abundance. 
Thus it is important to check the validity of the present laboratory constraints 
on the scale of \ncy at a macroscopic level.

The ``relevant set'' of particles during nucleosynthesis could be $u,~d$ quarks 
and electrons. Summation over spin of the particles is essentially zero,
however, the spin independent part gives a non-vanishing magnetic field, which
is given by
\be\label{mag}
\vec{B}=
2~\frac{e  \alpha \gamma_{\rm e}}{6\pi}\times
\vec{\theta}\times \sum_{i}(q_i^3 m_{i})\ n_i\,.
\ee
Considering the net contribution of nucleons (proton and
neutron) by their constituents $u,~d$ quarks, and taking the masses
of these quarks to be $\sim 5 {\rm MeV}$, we can estimate the net magnetic field,
which yields
\be
\label{esti}
B \approx (10^{-7}~-~10^{-8})\times \theta \times n_{\rm B}~{\rm GeV^{2}} \,,
\ee
where for rough estimation we have replaced $n_{i}$ by the baryon number
density $n_{\rm B}$ at the time of nucleosynthesis. Knowing that a successful 
nucleosynthesis requires $n_{\rm B}/n_{\gamma} \approx 10^{-10}$, where
$n_{\gamma}= (2\zeta(3)/\pi^2){\rm T}^3$ is the photon number density. Taking all
these into account along with the scale of \ncy $\Lambda_{\rm NC} \sim 10^{3}$GeV, we
get a magnetic field strength during nucleosynthesis ${\rm B}({\rm T}=10{\rm MeV}) 
\sim 10^{-31} {\rm GeV}^2$, which is much weaker to cause any kind of threat 
to nucleosynthesis. Thus, the global presence of \ncy is hardly felt during 
nucleosynthesis. In fact similar calculation can be repeated at later stages of 
the evolution also.

On can repeat similar analysis during the galaxy formation at temperature
${\rm T}_{\rm gf} \approx 10^{-3}{\rm eV}$, and, we get the number density of baryons in
the horizon, $n_{\rm B} \approx 10^{-41}{\rm GeV}^{3}$, which leads to net magnetic
field with a strength ${\rm B}_{\rm gf} \approx 10^{-34}$G. This is again weaker 
compared to $10^{-20} -10^{-30}$G, required for the amplification by the dynamo 
process to the present magnetic field in the galaxies. This again illustrates
that perhaps the presence of \ncy can hardly influence the global structure 
of the Universe. It can be easily verified that the present contribution 
to the magnetic field due to \ncy is again much smaller than $10^{-9}$ G, a lower
bound from COBE observation which constraints the anisotropic stresses due to 
the presence of magnetic field.  The reason behind such a weak effect is 
due to the fact that $\theta \propto 1/\Lambda_{\rm NC}^2$, any enhancement 
in the \ncy scale makes any observable global affect weaker in magnitude. 
For instance, for a given number density of charged massive particles the 
magnetic field is inversely proportional to $\Lambda^2_{\rm NC}$, and, 
its contribution to energy momentum tensor is suppressed by $\Lambda^4_{\rm NC}$.
However, things might change if we go beyond electroweak scale. This we 
discuss in the rest of this paper.

So far we have limited our discussion below 
electro-weak scale, during electro-weak scale there will be a little  enhancement 
in the magnetic field due to higher temperature and mass of the top quark which 
will dominate the particle spectrum. For $\Lambda_{\rm NC} = 10^{3}$GeV, the strength
of the magnetic field comes around ${\rm B}_{\rm ew}\sim 10^{-17}{\rm GeV}^2$. 
If magnetic flux conservation holds good so that the energy contribution due 
to the magnetic field acts as a source of radiation, then, the lines of magnetic fields
are frozen along with the expansion of the Universe and follow 
${\rm B}\propto {\rm T}^2$\footnote{which need not be guaranteed in presence of 
\ncy. More precisely, NC$U(1)$ theory, unlike its \com counter-part, is
an interacting theory, i.e. in the action and therefore in Hamiltonian
besides the terms quadratic in fields we have some terms of power three
and four. This will change the ${\rm T}^2$ behaviour at temperatures higher
than \ncy scale.}. Therefore the magnetic field produced during the 
electroweak scale will have a strength $\sim 10^{-25}{\rm GeV}^2$ during 
nucleosynthesis. This number is at least six orders of magnitude better than 
the magnetic field produced due to \ncy during nucleosynthesis.
This leads to an obvious suspicion that at higher temperatures perhaps  
we might be able to get some appreciable affects due to \ncy. However,
to proceed with this we need to speculate an exact strength of the magnetic field.

At temperatures more than the electroweak scale (before the electroweak
phase transition) all the fermions are massless, so they do not contribute
to large scale magnetic fields. In the usual Higgs scenario the Higgs
field is the only massive field before the electroweak phase transition.
It may seem that, being neutral, Higgs is not entering into the game.
However, one should note that the fluctuations of the upper component in
the Higgs doublet will lead to a charged massive particle. These particles
after the electroweak phase transition are absorbed in the zero helicity
part of the massive gauge bosons, $W^{\pm}$. So, we have fulfilled the
required assumption of  having massive charged particles. However, we note that
Eq.~(\ref{magnet}) is for fermionic field. For the
scalar fields, such as Higgs, we believe that we  still maintain the
structural form of Eq.~(\ref{magnet}), but with a different numeric
factor. However, Eq.~(\ref{magnet}) will still serve our purpose for an order
of magnitude calculation. So, for a rough estimation we consider the magnetic
field is given by
\be\label{B_H}
B \approx (10^{-3}~-~10^{-4})\times {1\over \ncs^2} \times m_{\rm H}\ n_{\rm H}\ ,
\ee
where $m_{\rm H}$, $n_{\rm H}$ are the Higgs mass and number density, respectively. 
It is worth noting that since the running of the coupling constant ($\alpha$) 
with temperature is logarithmic, this will not change our analysis appreciably.
In order to get the temperature dependent $m_{\rm H}$, we use the usual Higgs model 
(e.g. \cite{Weinberg})
\be\label{Higgs}
V(\Phi) = -{\mu \over 2} ~\left[ \left({{\rm T} \over {\rm T}_{\rm ew}}\right)^2 -
1 ~\right] \Phi^2 + {\lambda \over 4} \Phi^4\ ,
\ee
where $\Phi$ is the Higgs doublet, ${\rm T}_{\rm ew}$ is the electroweak phase 
transition temperature, and the usual electroweak scale and data (like Higgs mass) is
defined by a proper choice of the constants $\mu$ and $\lambda$. From Eq.~(\ref{Higgs}),
one can read the temperature dependent Higgs mass as a coefficient of the first term.
For ${\rm T}\gg {\rm T}_{\rm ew}$, $m_{\rm H}\propto {\rm T}$, and $n_{\rm H}
\propto {\rm T}^3$. Since, the magnetic field $B\propto T^2$, one can easily 
read off the temperature dependence of the $\ncs$. In a more general discussion 
$m_{\rm H}\ n_{\rm H}$ in Eq.~(\ref{B_H}) can be
replaced by the energy density of the massive charged  particles, $\rho$.
For a relativistic scalar particles $\rho \propto T^4$, which gives 
a simple expression for $\ncs$ beyond electroweak scale
\be
\ncs \simgt \epsilon T\,,
\ee
where $\epsilon $ being a constant which in our Higgs model 
is of the order of $\epsilon \approx  {\cal O}(10 -100)$. The 
upper value of $\epsilon$ is fixed by demanding that the strength of the 
magnetic field is constrained via nucleosynthesis. However, 
the reader might have noticed that below the electroweak scale 
the temperature dependence of $\ncs$ is $\propto {\rm T}^{1/2}$, because 
the essential charged particles are baryons whose number density scales as 
$a^{-3}$, where $a$ is the cosmological scale factor.   
The temperature dependence of \ncy has also been conjectured in Ref.\cite{Chu}, and 
some cosmological consequences have been discussed. However, in thier case they just
assume this temperature dependence to be step-like. 
       
Finally on a speculative note we mention that it is possible to have 
GUT baryogenesis at a very high temperature in an inflationary model. A temperature 
which can be at least $2-3$ orders of magnitude larger than the usual reheat 
temperature of the Universe \cite{chung}. Usually, the reheat temperature is bounded by 
gravitino over production during reheating and it is usually set to be less than 
$10^{9}$GeV. In that case it is possible to produce a large magnetic field due to 
\ncy. If such models are taken seriously then it is also possible to constrain 
$\ncs$ from the upper bound on strength of the magnetic field during nucleosynthesis.
However, due to many cosmological reasons a low reheat temperature is favourable and 
that is why we do not delve into the details of constraining $\ncs$ from the very 
early Universe.

We summarize our paper by mentioning that the presence of \ncy at a scale $\ncs \approx
10^{3}$GeV does not cause any serious problem at the cosmological scale. Considering
the \ncy as a seed for the primordial magnetic field, we have studied the temperature
dependence of $\ncs$. We have shown that this temperature dependence goes as
$T^{1/2}$ for $T\simlt T_{{\rm ew}}$ and for $T\gg T_{{\rm ew}}$, $\ncs \propto T$, 
where $T_{{\rm ew}}$ is the electro-weak phase transition temperature.
Although the magnetic field we have found  for $\ncs \approx 10^{3}$ is less than the
required seed for the usual dynamo effect, it can align the particles spin so that they
can result in a net magnetic field providing the seed \cite{prog}.

However, the present constraints on temperature of the Universe beyond nucleosynthesis
is so weakly constrained that we can not specifically define the \ncy scale $\ncs$
without invoking a particular model of high or low reheat temperatures. If we 
choose inflationary models with a high reheat temperature then it is quite possible
that \ncy will play an interesting role during that period.

\vskip 0.5cm

{\bf Acknowledgments} 

This work is partially 
supported by the EC contract no. ERBFMRX-CT 96-0090, and A. M. acknowledges 
the support of {\bf The Early Universe network} HPRN-CT-2000-00152.


\end{document}